\documentclass[pre,amsmath,amssymb,superscriptaddress,showpacs,reprint]{revtex4-1}

\usepackage{amsmath,amssymb,graphicx}
\usepackage{amsbsy}
\usepackage{latexsym}
\usepackage{color}
\usepackage{graphicx}
\usepackage{psfrag}
\usepackage[normalem]{ulem}
\usepackage{bm}

\newcommand{\be}{\begin{equation}}
\newcommand{\ee}{\end{equation}}
\newcommand{\bea}{\begin{eqnarray}}
\newcommand{\eea}{\end{eqnarray}}

\newcommand{\comment}[1]{}

\begin{document}

\title{Asymptotic States of Ising Ferromagnets with Long-range Interactions}

\author{Ramgopal Agrawal}
\email{ramgopal.sps@gmail.com}
\affiliation{School of Physical Sciences, Jawaharlal Nehru University, New Delhi 110067, India.}
\author{Federico Corberi}
\email{corberi@sa.infn.it}
\affiliation{Dipartimento di Fisica ``E.~R. Caianiello'', and INFN, Gruppo Collegato di Salerno,
and CNISM, Unit\`a di Salerno, Universit\`a  di Salerno, via Giovanni Paolo II 132, 84084 Fisciano (SA), Italy.}
\author{Ferdinando Insalata}
\email{ferdinsa@live.it}
\affiliation{Department of Mathematics, Imperial College London, London SW7 2AZ, United Kingdom.}
\author{Sanjay Puri}
\email{purijnu@gmail.com}
\affiliation{School of Physical Sciences, Jawaharlal Nehru University, New Delhi 110067, India.}

\begin{abstract}
It is known that, after a quench to zero temperature ($T=0$), two-dimensional ($d=2$) Ising ferromagnets with short-range interactions do not always relax to the ordered state. They can also fall in \textit{infinitely} long-lived striped \textit{metastable} states with a finite probability. In this paper, we study how the abundance of striped states is affected by long-range interactions. We investigate the relaxation of $d=2$ Ising ferromagnets with power-law interactions by means of Monte Carlo simulations at both $T=0$ and $T \ne 0$. For $T=0$ and the finite system size, the striped metastable states are suppressed by long-range interactions. In the thermodynamic limit, their occurrence probabilities are consistent with the short-range case. For $T \ne 0$, the final state is always ordered. Further, the equilibration occurs at earlier times with an increase in the strength of the interactions.
\end{abstract}

\maketitle

\section{Introduction}
\label{S1}

The domain coarsening in ferromagnetic systems following a quench below the critical temperature is a well-understood phenomenon~\cite{PuriWad09,Bray94,CRP,Cor15,CorCugYos11}, whereby the locally magnetized domains of competing phases grow in time. For finite systems, such competition comes to an end at time $t_m$ when the typical size of growing domains $R(t)$ becomes comparable with the system size $L$. At this point, the phase symmetry is broken, and the macroscopic domains of the single phase appear. This relaxation to the final state can take different routes. One naive expectation is that a single large domain may start prevailing at times $t\simeq t_m$ and quickly invades the whole system. On the other hand, it is also possible that the system may reach some \textit{metastable} configuration. Depending on the stability of interfaces in such a configuration, system can get trapped forever at zero temperature ($T=0$). The thermal activation is then required to escape by removing remaining interfaces.

\begin{figure}[b]
	\centering
	\includegraphics[width=0.42\textwidth]{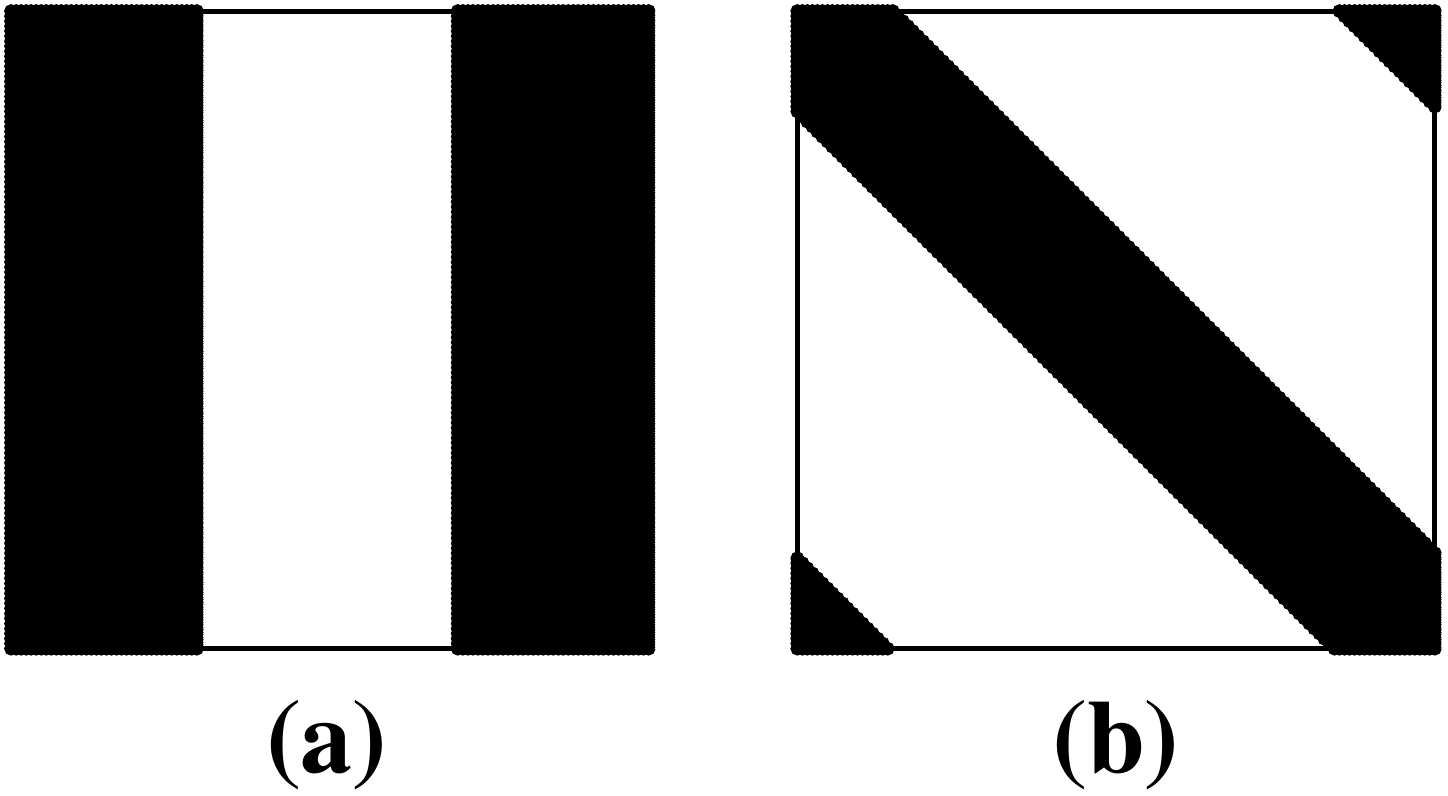}
	\caption{The typical metastable states on a $2d$ square lattice with periodic boundary conditions. (a) Stripes in vertical direction. (b) Stripes in diagonal direction.}
	\label{fig1}
\end{figure}

In the last two decades, there has been a great interest in zero-temperature relaxation of Ising ferromagnets. In dimension $d=1$, they always reach the ordered state due to trivial geometry, while in $d \ge 3$, they never reach the ordered state~\cite{PhysRevE.63.036118,PhysRevE.65.016119,PhysRevE.83.030104,*PhysRevE.83.051104}. The $2d$ Ising ferromagnets are even more intriguing and show a surprising connection with percolation theory~\cite{stauffer2018introduction}. Previous studies~\cite{BarKraRed09,PhysRevLett.109.195702,BlaPic13} revealed that on the square lattice with short-range interactions, a fraction $\pi_+\simeq 0.62$ of the realizations reach the ordered state, while fraction $\pi_-\simeq 0.34$ of the realizations reach the \textit{infinitely} long-lived metastable state with vertical or horizontal stripes (Fig.~\ref{fig1}(a)). Perhaps a small fraction $\pi_/\simeq 0.04$ of the realizations can reach a metastable state with diagonal winding stripes (Fig.~\ref{fig1}(b)). The latter slowly decay to the completely ordered state on a time-scale $t_{diag} \sim L^3 \gg t_m$~\cite{PhysRevE.63.036118}. The probabilities ($\pi_+, \pi_-, \pi_/$) of reaching distinct states are identical to the crossing probabilities of clusters in the $2d$ random site percolation model at percolation threshold~\cite{PhysRevLett.109.195702,Pinson94}.

It was recently realized~\cite{BlaCorCugPic14} that the choice among the above possibilities is taken at a time $t_p$ occurring much before $t_m$. More precisely, despite $t_p$ and $t_m$ diverging when $L\to \infty$, one has $\lim _{L\to \infty}t_p/t_m=0$. At such early times, the connected domains do not extend too far, and on large scales, the physics is mostly controlled by a fast-growing \textit{critical percolation} structure touching the boundaries of the system at $t=t_p$. The way this percolation structure crosses the system (spanning in both lattice directions, only along one of them, or spanning in diagonal directions) fully determines, as early as at $t=t_p$, the final destiny occurring at $t>t_m$. This physics has been shown to be quite general, at least in $2d$, being robust against different lattice choices~\cite{Blanchard_2017} and presence of quenched disorder~\cite{CorCugInsPic17,CorCugInsPic19,InsCorCugPic18}. For low but non-zero $T$ quenches, the early time behavior of the system is governed by the $T=0$ fixed point. Thus, the above scenario applies to these cases as well. Only at large times the convergence to an ordered state takes place.

An important and unexplored question is how this phenomenology is influenced by the presence of \textit{long-range} interactions. Considering an interaction decaying as $1/r^{d+\sigma}$, a first important distinction can be made between the cases $\sigma \le 0$ and $\sigma > 0$. In the first case, sometimes denoted as the strong long-range regime, some fundamental physical properties, such as \textit{additivity}, are spoiled, causing major differences with respect to the kinetics of the short-range case. In particular, it was shown~\cite{CorIanKumLipPol21} that, at least in $d=1$, a mean-field  like dynamics dominates in the thermodynamic limit $L\to \infty$. In this case, spins tend to align with the sign of global magnetization, the symmetry is soon broken, and metastable states such as those described before for short-range systems cannot be sustained. For $\sigma >0$, instead, in the so-called weak long-range regime, one observes a coarsening phenomenon qualitatively similar to the one of short-range systems. Despite that, it is reported in recent works~\cite{PhysRevE.103.012108,PhysRevE.103.052122} that the striped metastable states (typical of those systems) are absent or significantly suppressed in the presence of long-range interactions. This can be understood at a qualitative level because the extended interaction allows single spins to probe the local magnetization far away, and this may provide a global drift to their update, even if it is not as efficient as in the mean-field. However, our understanding of the phenomenon does not go much beyond this vague intuition, and several questions remain open, among which if metastability is fully suppressed or only penalized by the long-range couplings.

In this paper, we study quantitatively and in detail this problem by means of numerical simulations of the $2d$ Ising model on the square lattice. In doing that, we uncover a quite rich scenario where the abundance of metastable states and their duration is determined by the interplay between the exponent $\sigma$ and system size $L$.

This paper is structured as follows. In Sec.~\ref{S2}, we introduce the model and provide simulation details. In Sec.~\ref{S3}, we present our main simulation results, i.e., for a quench to zero temperature. We also infer some scaling laws to describe the data. In Sec.~\ref{S4}, we discuss the case of non-zero temperature quenches. In Sec.~\ref{S5}, we summarize the results obtained in this paper. In Appendices~\ref{A1} and \ref{A2}, some additional results for $\sigma = 0.6$ and $\sigma = 1.5$ are provided. Appendix~\ref{A3} contains results for the pair connectedness function, which is usually considered in percolation theory.

\section{Model and Simulation details}
\label{S2}

We consider the ferromagnetic Ising model with long-range couplings, whose Hamiltonian is given as
\be
{\cal H}(\{s_i\})=- \sum_{i < j} J(r) s_i s_j,
\label{ham}
\ee
where $s_i=\pm 1$ are Ising spins on sites $i$ of the two-dimensional square lattice of linear size $L$ with periodic boundary conditions (PBC). Spins $s_i, s_j$ at distance $r=\vert i -j \vert$ interact with ferromagnetic coupling constant
\be
J(r)=\frac{1}{r^{(2+\sigma)}}.
\ee
Letting $J(r)=\delta_{r,1}$, one recovers the usual nearest neighbor (NN) model. In equilibrium, this model has a \textit{para-ferromagnetic} phase transition at a finite critical temperature $T_c(\sigma)$~\cite{PhysRevB.1.2265,Fisher1972critical,PhysRevB.8.281,PhysRevB.56.8945,PhysRevLett.89.025703,PhysRevE.95.012143}.

The single spins flip with Metropolis transition rates $L^{-2} \min [1,\exp(-\Delta E/T)]$, with the Boltzmann constant set to unity. Here $\Delta E$ is the energy difference in the proposed move. Time is measured in Monte Carlo steps (MCS), each corresponding to $L^2$ attempts of spin flips. Initially, the system is in equilibrium at infinite temperature, which is suddenly quenched to $T<T_c(\sigma)$ at the time $t=0$ and afterward evolved at this temperature with Metropolis transition rates. The Ewald summation technique~\cite{PhysRevE.103.012108,fs2002,PhysRevE.95.012143} is used to handle the pairwise long-range interaction between any two spins on the periodic lattice. To assure the accuracy of the data, the number of initial configurations used for each $L$ is typically $10^4-10^6$.

According to the Bray-Rutenberg predictions~\cite{BrayRut94,CMJ19}, for a quench to non-zero $T$, the model in Eq.~\eqref{ham} exhibits the domain growth law $R(t)\sim t^{1/z}$. The exponent $z=2$ for $\sigma >1$ and $z=1+\sigma$ for $\sigma \le 1$, with logarithmic corrections at $\sigma =1$. For quenches to $T = 0$, a different exponent ($z = 4/3$) has been observed for all $\sigma$~\cite{PhysRevE.103.012108,PhysRevE.103.052122}. At small $T$, a crossover from this universal behavior to the Bray-Rutenberg law is observed~\cite{PhysRevE.103.012108}.

Here, we evaluate the average size $R(t)$ of growing domains as the inverse excess defect density, $R(t)\equiv [\rho (t) -\rho _{eq}]^{-1}$, $\rho (t)$ being the fraction of anti-aligned spins and $\rho _{eq}$ its equilibrium value~\cite{Bray94}.

\begin{figure}[t]
\centering
	\includegraphics[width=0.49\textwidth]{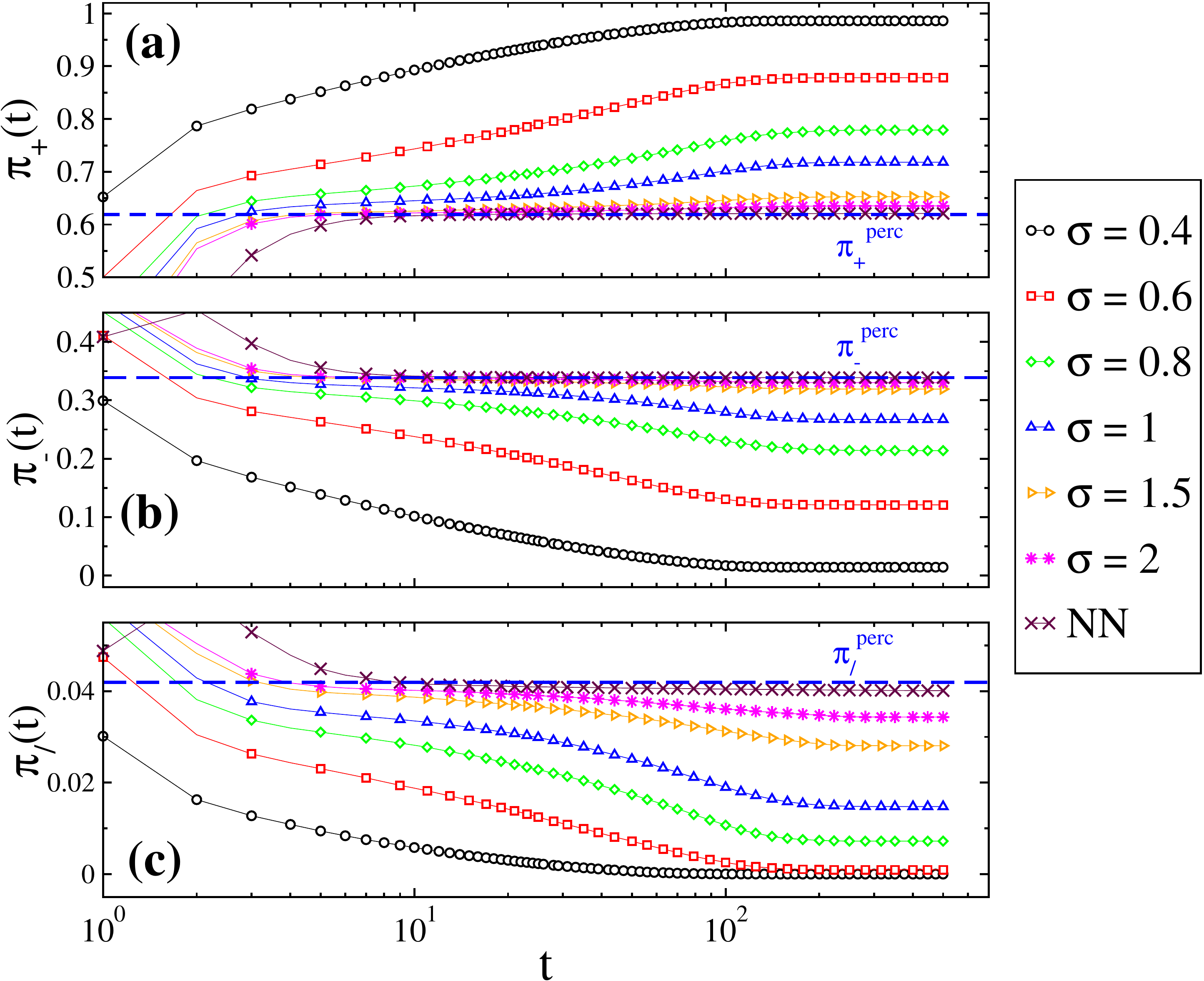}
	\caption{The crossing probabilities (a) $\pi_+(t)$, (b) $\pi_-(t)$, and (c) $\pi_/(t)$ are plotted against $t$ (with log-linear scales) for different values of $\sigma$ (see key), for a quench to $T=0$ of a system of linear size $L=128$. The horizontal blue dashed lines represent the values of $\pi_+^{perc}$, $\pi_-^{perc}$,
		$\pi_/^{perc}$.}
	\label{fig2}
\end{figure}

\begin{figure}[t]
	\centering
	\includegraphics[width=0.48\textwidth]{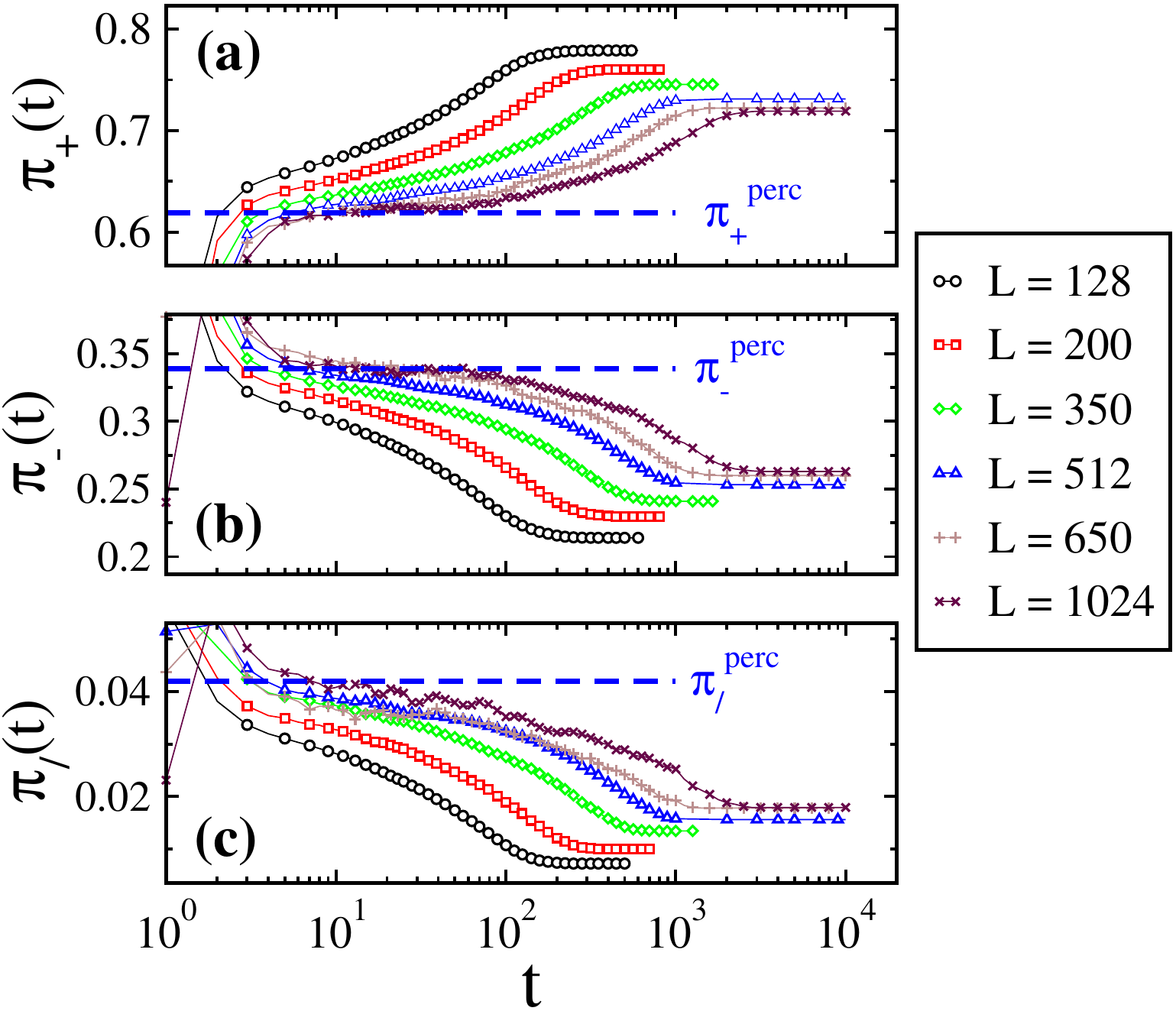}
	\caption{The crossing probabilities (a) $\pi_+(t)$, (b) $\pi_-(t)$, and (c) $\pi_/(t)$ are plotted against $t$ (with log-linear scales) for a quench to $T=0$ with $\sigma=0.8$ for systems of different sizes $L$, see key. The horizontal blue dashed lines represent the values of $\pi_+^{perc}$, $\pi_-^{perc}$, $\pi_/^{perc}$.}
	\label{fig3}
\end{figure}

\section{Main results}
\label{S3}

An important quantity to investigate the relaxation process is the time-dependent crossing probability $\pi(t)$. It is the probability that a connected cluster of parallel spins spans the system at time $t$. Such cluster can cross the lattice from one side to the other horizontally or vertically. We denote the probability of these events as $\pi _-(t)$. The clusters can also percolate along one of the two diagonal directions with equal probability $\pi _/(t)$. Alternatively, a cluster can traverse the system in both horizontal and vertical directions, with probability $\pi_+(t)$. To compact the notation, these quantities will also be generically denoted as $\pi _x(t)$, with $x=+,-,/$. With NN interactions the long time values of these probabilities equal the corresponding (time-independent) values computed~\cite{Pinson94} in two-dimensional critical percolation, namely, $\pi_+^{perc}=0.61908$, $\pi _-^{perc}=0.3388$, and $\pi _/^{perc}=0.04196$. With PBCs, wrapping the torus more than once is also possible, but the associated probability is very small, and we will not keep track of these events in this study.

In Fig.~\ref{fig2}, we plot the crossing probabilities $\pi_x(t)$ against $t$ for a system of size $L=128$ quenched to $T=0$, for different values of $\sigma$. The NN case is also shown for comparison. The figure shows that, for any value of $\sigma$, the crossing probabilities are monotonous and display a two-step behavior. They quickly attain a pre-asymptotic plateau of height $\pi _x^{early}$. This sets in around $R(t)\simeq 5$. Later on, there is an evolution towards another plateau at $\pi_x=\pi_x^{late}$. The same figure shows that, for small $\sigma$, $\pi_x^{early}$ overshoot the percolation values $\pi_x^{perc}$, while for large $\sigma$ they approach $\pi_x^{perc}$. The late time values $\pi_x^{late}$ also show deviation from $\pi_x^{perc}$ and approach towards the ordered state ($\pi_-=\pi_/=0$, $\pi_+=1$) when lowering $\sigma$. For larger $\sigma$, the evolution completes at early times with $\pi_x^{late}$ ending up close to $\pi_x^{perc}$. One might expect that, around a critical value of $\sigma$ (in the equilibrium context), a qualitative change between weak long-range and short-range behavior may occur in Fig.~\ref{fig2}. Following the literature~\cite{PhysRevE.95.012143,PhysRevB.8.281}, $\sigma = 1$ is the lower critical decay exponent below which the critical exponents match with the mean-field values. Further, $\sigma = 7/4$ is the upper critical decay exponent above which the critical exponents are those of the NN model. From the data in Fig.~\ref{fig2}, nothing special is seen to happen around $\sigma = 1$. However, for $\sigma \gtrsim 2$ the data shows behavior akin to the NN case. From the numerical data, one does not observe a sharp crossover at the critical values of $\sigma$. Notice that in the NN case, as it is known, the percolation values $\pi_x^{perc}$ are perfectly attained in Fig.~\ref{fig2}. This data of small system ($L=128$) stresses that, for smaller $\sigma$, the asymptotic striped states are suppressed. However, as we will see below, the system size plays a major role for long-range interactions, and the data from small systems is insufficient to make general claims.

We investigate the role of the system size via repeating the same kind of simulations described above upon varying $L$. In Fig.~\ref{fig3}, we show results for $\sigma = 0.8$ (see Appendix~\ref{A1} for $\sigma = 0.6$ and $1.5$). We see that, with an increase in $L$, $\pi _x^{early}$ quickly set to $\pi_x^{perc}$ (similar behavior is found for other values of $\sigma$). It indicates that for any value of $\sigma$, the long-range model builds at short times a critical percolation structure. We will shortly probe this behavior in detail. Furthermore, increasing $L$ has another systematic effect: the late saturation values $\pi_x^{late}(L)$ also approach the percolation values $\pi_x^{perc}$ and the time they reach increases with $L$. This evolution of $\pi_x(t)$ beyond the first plateau at $\pi_x^{\rm early}$ to $\pi_x^{late}$ can be understood in terms of the following scaling law:
\be
\label{newscale}
\pi_x(t,L)= \pi_x^{perc} + \Delta_x (L) f_x\left[\frac{R(t)}{L}\right] .
\ee
The argument of the scaling function $f_x(z)$ cannot exceed $1$ as $R(t) \le L$. The limiting behaviors are $f_x(z) \rightarrow 0$ for $z \ll 1$ and $f_x(z) = 1$ for $z \simeq 1$. The quantities $\Delta_x (L) \left[= \pi_x^{late}(L)-\pi_x^{perc}\right]$ are the overshoot from the percolation values $\pi_x^{perc}$ (positive for $\pi_+$, and negative for both $\pi_-$ and $\pi_/$).

\begin{figure}[t]
	\centering
	\includegraphics[width=0.38\textwidth]{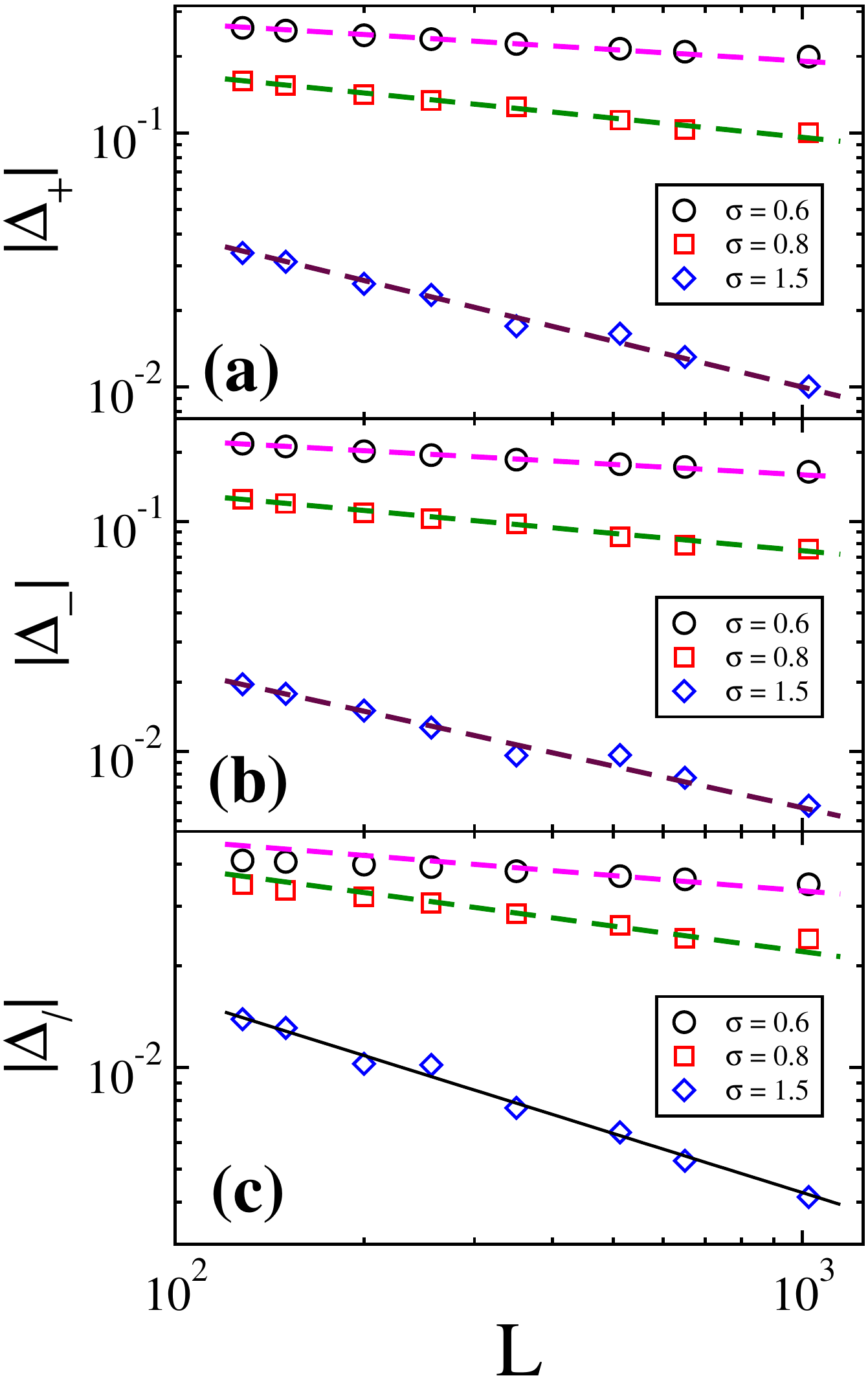}
	\caption{The absolute values (a) $\vert \Delta _+(L) \vert$, (b) $\vert \Delta _-(L) \vert$, and (c) $\vert \Delta _/(L) \vert$ are plotted against $L$ with double logarithmic scales, for $T = 0$ quench with $\sigma =0.6$, $\sigma=0.8$ and $\sigma=1.5$. The straight dashed lines indicate the behavior $L^{-\alpha}$ with $\alpha=0.15$ for $\sigma=0.6$, $\alpha=0.25$ for $\sigma=0.8$ and $\alpha=0.6$ for $\sigma=1.5$. The solid line in (c) denotes the best fit: $\vert \Delta_/(L) \vert \simeq 0.23 \times L^{-(0.58 \pm 0.03)}$.}
	\label{fig4}
\end{figure}

To sensitively check the approach of $\pi_x^{late}$ to $\pi_x^{perc}$, we extract $\Delta_x$ from the saturation values of $\pi_x(t)$ for different $\sigma$ and plot their absolute values in Fig.~\ref{fig4}. They seem to show an algebraic behavior of the form
\be
\vert \Delta_x(L) \vert \propto L^{-\alpha}.
\label{eqDelta}
\ee
The value of $\alpha$ seems unique for the three quantities $\Delta _x$ and is dependent on $\sigma$: $\alpha \simeq 0.15$ for $\sigma =0.6$, $\alpha \simeq 0.25$ for $\sigma=0.8$, and $\alpha \simeq 0.6$ for $\sigma=1.5$. For $\sigma =0.6$, however, we find the same value $\alpha \simeq 0.15$ for $\Delta_+$ and $\Delta _-$, and somewhat smaller value $\alpha \simeq 0.08-0.09$ for $\Delta _/$. It could be due to the noisy character of the quantity $\Delta_/$ and possibly, to finite size effects, which are more pronounced upon lowering $\sigma$. Notice that $\alpha$ decreases with $\sigma$, indicating that the approach to the large-$L$ behavior is slower. In panel (c), the changes in $\vert \Delta_/(L) \vert$ with system size $L$ are very small for $\sigma = 0.6$ and $0.8$. Therefore, a much larger range of sizes would be needed, which is not currently feasible given the difficulty with handling long-range interactions. For $\sigma = 1.5$, on the other hand, $\vert \Delta_/(L) \vert$ shows a reasonable change, and the best fit yields $\alpha \simeq 0.58 \pm 0.03$, which is consistent with $\alpha \simeq 0.6$. Notice that, given the relatively small range of $L$ which we could access numerically, the scaling behavior in \eqref{eqDelta} is not compelling. This, however, has no consequences on the following analysis where the numerical values of $\Delta _x(L)$ [not the proposed form~(\ref{eqDelta})] will be directly considered.

\begin{figure}[t]
\centering
	\includegraphics[width=0.48\textwidth]{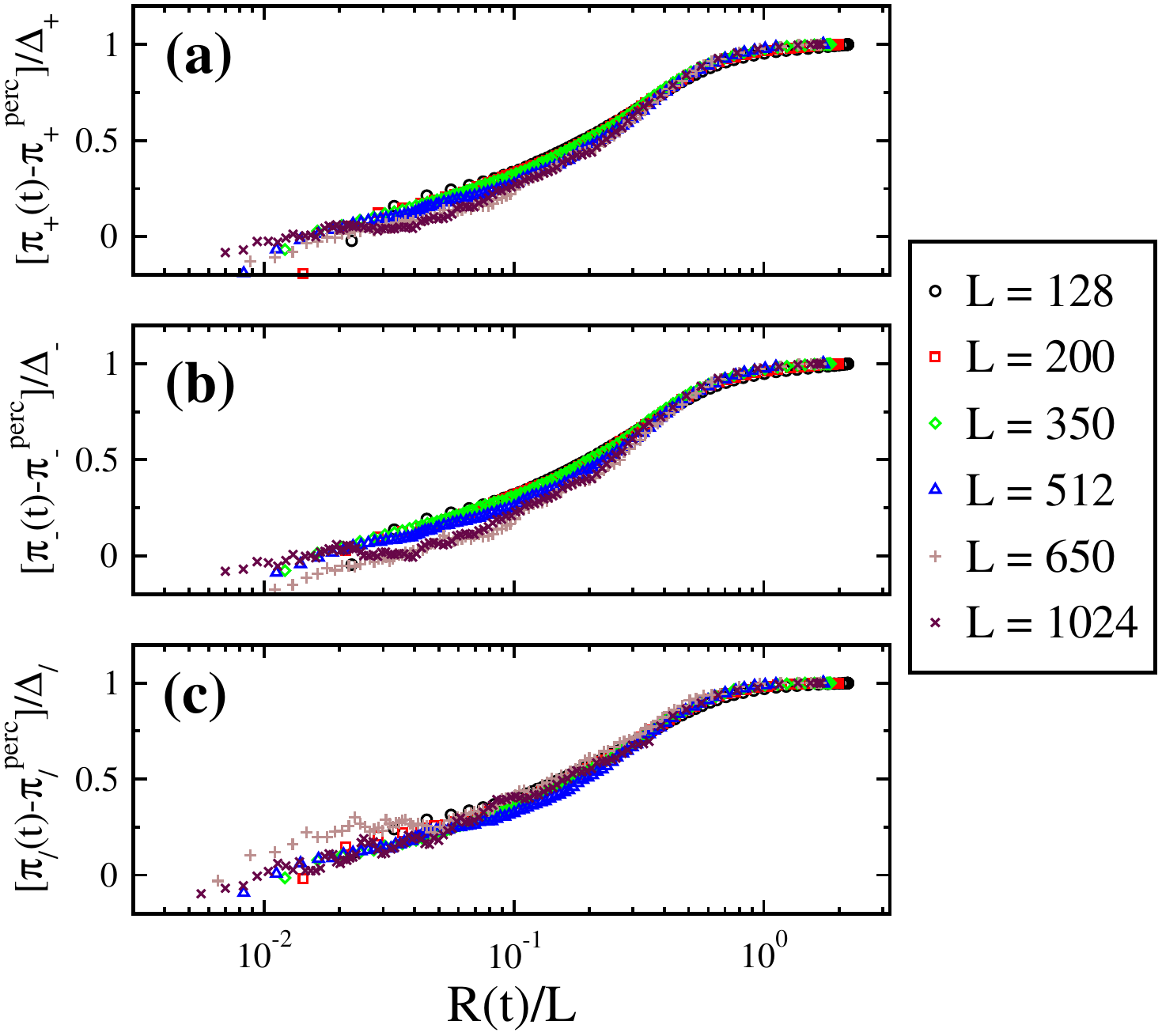}
	\caption{The quantities (a) $[\pi_+(t)-\pi_+^{perc}]/\Delta _+(L)$, (b) $[\pi_-(t)-\pi_-^{perc}]/\Delta _-(L)$, and (c) $[\pi_/(t)-\pi_/^{perc}]/\Delta _/(L)$ are plotted against $R(t)/L$ (with log-linear scales) for a quench with $\sigma=0.8$ to $T=0$ for systems of different sizes $L$.}
	\label{fig5}
\end{figure}

If the scaling in Eq.~\eqref{newscale} works, one should observe a data collapse when plotting $[\pi_x(t)-\pi_x^{perc}]/\Delta _x(L)$ vs. $R(t)/L$. This is demonstrated in Fig.~\ref{fig5} for $\sigma = 0.8$. For all three crossing probabilities, the observed data collapse is very good. In fact, we found that the data in panels (a)-(c) fall on the same scaling function (not shown here). Thus, the finite-size-scaling functions $f_x$ in Eq.~\eqref{newscale} are universal with respect to the different crossings. A similar behavior is observed for $\sigma = 0.6, 1.5$, as shown in Appendix~\ref{A2}. Note that the scaling proposed in Eq.~\eqref{newscale} is obeyed over the entire evolution of $\pi_x(t)$ after the first plateau.

We next turn to our observation regarding the critical percolation structure near the first plateau at $\pi _x^{early}$ and investigate the geometry of the clusters in the system. Note that, at critical percolation, the geometrical features in the system are \textit{fractal} objects. Therefore, when $\pi_x(t) \simeq \pi_x^{perc}$, the connected spin clusters of the long-range system should have the fractality of the critical percolation class. The (average) squared winding angle $\langle \theta ^2(r)\rangle$ is one of the quantities to test this behavior sensitively. For geometrical spin clusters this quantity is calculated as follows: At a given time, two points $i,j$ are chosen on the external perimeter --- the hull --- of a cluster and the winding angle $\theta _{ij}$, namely the angle (measured counterclockwise) between the tangents to the perimeter at $i$ and $j$, is computed. The procedure is repeated for all the couples of perimeter points at distance $r$ measured along the hull, taking square, and averaging over such measurements is denoted by $\theta^2(r)$. Further, averaging $\theta ^2(r)$ over the non-equilibrium ensemble, one ends up with $\langle \theta ^2(r)\rangle$. This quantity is known exactly in the $2d$ random percolation at percolation threshold $p = p_c$~\cite{PhysRevLett.60.2343,PhysRevE.68.056101}, and is given as
\be
\langle \theta ^2 (r,r_0)\rangle = a +\frac{4k}{8+k}\ln \left (\frac{r}{r_0}\right ),~~~~r\gg r_0
\label{angleperc}
\ee
with $k = 6$, a number related to the fractal dimensions of cluster area and cluster interface at critical percolation~\cite{Blanchard_2017}. Here $a$ is a non-universal constant and $r_0$ is the lattice spacing. When computing this quantity in the coarsening system, we consider only the largest cluster for numerical convenience.

\begin{figure}[t!]
	\centering
	\includegraphics[width=0.48\textwidth]{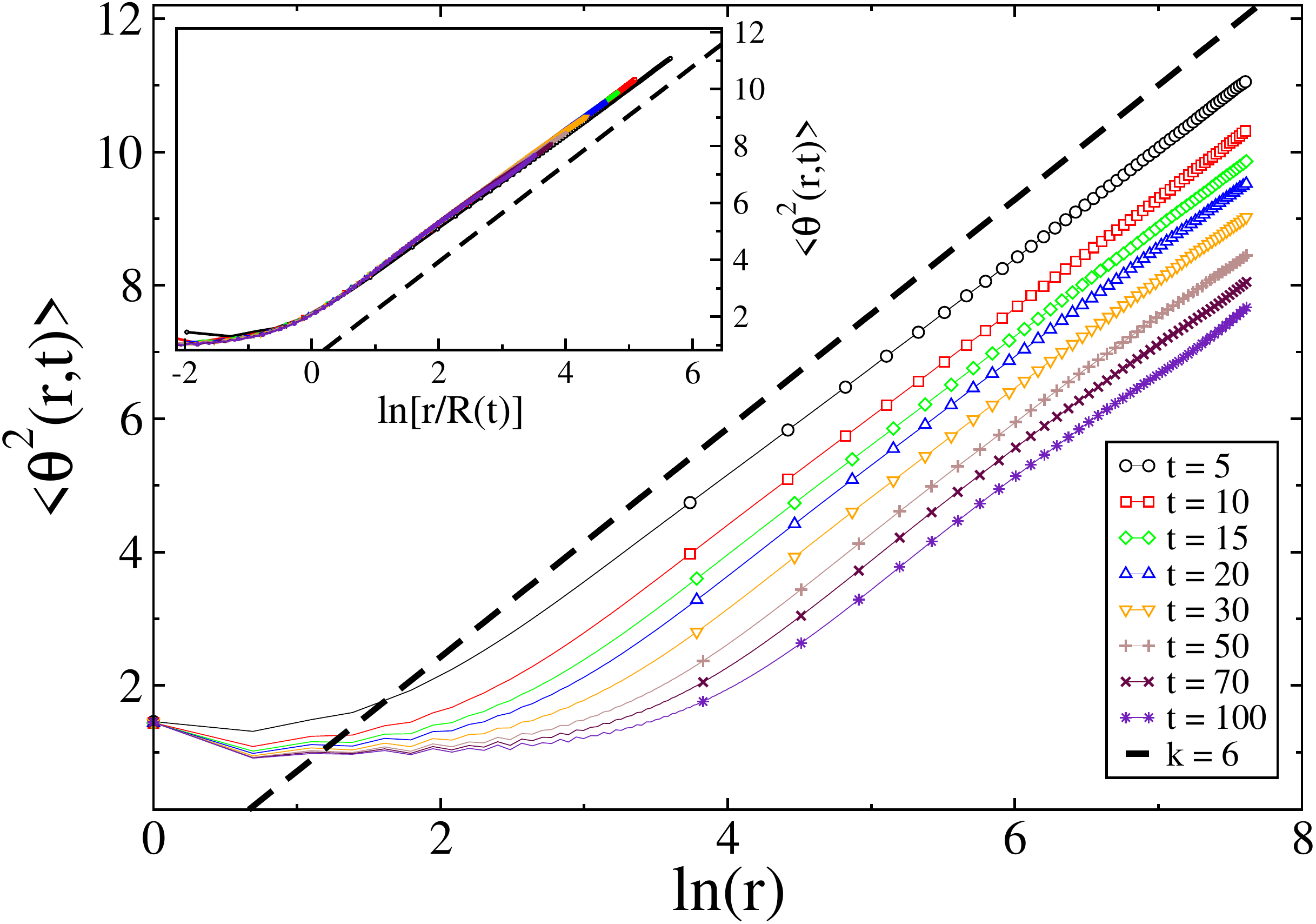}
	\caption{The average squared winding angle $\langle \theta ^2(r,t)\rangle$ is plotted against $\ln r$ for a quench 
		with $\sigma =0.8$ to $T=0$, in a system of size $L=1024$ with linear scales. In the inset, the same quantity is plotted
		against $\ln[r/R(t)]$. The dashed curves are straight lines with slope $k=6$, obtained from Eq.~\eqref{angleperc}.}
	\label{fig6}
\end{figure}

In the case with NN interactions~\cite{Blanchard_2017,CorCugInsPic17,CorCugInsPic19}, the law~(\ref{angleperc}) is obeyed with the replacement $r_0\to R(t)$. i.e., the growing structure in a coarsening system has the property of a critical percolation cluster on an {\it effective} lattice with spacing $R(t)$. With long-range interactions, we find the same result, as it can be seen in Fig.~\ref{fig6} for $\sigma = 0.8$. In the main frame of the figure, one sees that, up to a certain value of $r$ that increases with time, the curves at different times deviate from the slope $k=6$ of critical percolation. This value of $r$ is identified with $R(t)$. For large value of $r$ the curves increase linearly with $\ln(r)$, with the slope $k=6$. In the inset, a nice collapse is observed in terms of scaling variable $r/R(t)$, compatible with Eq.~(\ref{angleperc}). The different data sets considered in Fig.~\ref{fig6} are in the range of time whereby the crossing probabilities $\pi_x$ stay near $\pi_x^{perc}$. i.e., $\pi_x(t) \simeq \pi_x^{perc}$. Similar results were found for the other values of $\sigma $ considered, in the range $[0.6-3]$.

We also performed a careful investigation of the pair-connectedness function, another quantity to testify the critical percolation signature (see the Appendix~\ref{A3}). Also this quantity, as the winding angle, shows that the cluster geometry in long-range systems resembles that of critical percolation.

The above reasoning confirms that, similar to the NN case, a critical percolation structure is formed at time $t_p$ in the presence of long-range interactions. Further, the same structure remains in the system up to the longest times~\footnote{Due to the power-law fall of the interaction, for any $\sigma$, the flat stripes formed in vertical/horizontal directions (Fig.~\ref{fig1}(a)) as well as the stripes formed in diagonal directions (Fig.~\ref{fig1}(b)) remain infinitely long-lived at zero temperature.} in the thermodynamic limit, as clear from Eq.~\eqref{eqDelta}. Thus, for $L\to \infty$, there are basically no main differences between the long-range and the NN case. However, for any finite $L$, the extended interactions are able to convert a fraction $\Delta _-\,+\,\Delta_/$ of realizations spanning along one single direction into double spanning ones. Given Eq.~(\ref{eqDelta}), for even sufficiently large values of $L$, this fraction can be high, thus suppressing the striped states at long times. This explains the observation made in Refs.~\cite{PhysRevE.103.012108,PhysRevE.103.052122} regarding the substantial absence of metastable states: According to the present study, this is a \textit{mere} finite-size effect.

As to the physical origin of this behavior, a possible interpretation is the following. In the weak long-range regime considered here ($\sigma >0$), the interaction is integrable, i.e., $\sum_{\vec{r}} J(r) < \infty$. Hence, one expects that a magnetized part cannot influence regions sufficiently far away. Then, in the thermodynamic limit, competing regions with opposite magnetization are expected, a situation which results in the formation of striped states. However, if the system is small enough, a magnetized part will influence the whole system, leading to a single domain (no stripes). As the decay of the power-law interaction is scale-free, the crossover between the two behaviors is expected to be broad and regulated by algebraic forms, which is indeed observed.

\section{Role of Thermal fluctuations}
\label{S4}

\begin{figure}[t!]
	\centering
	\includegraphics[width=0.48\textwidth]{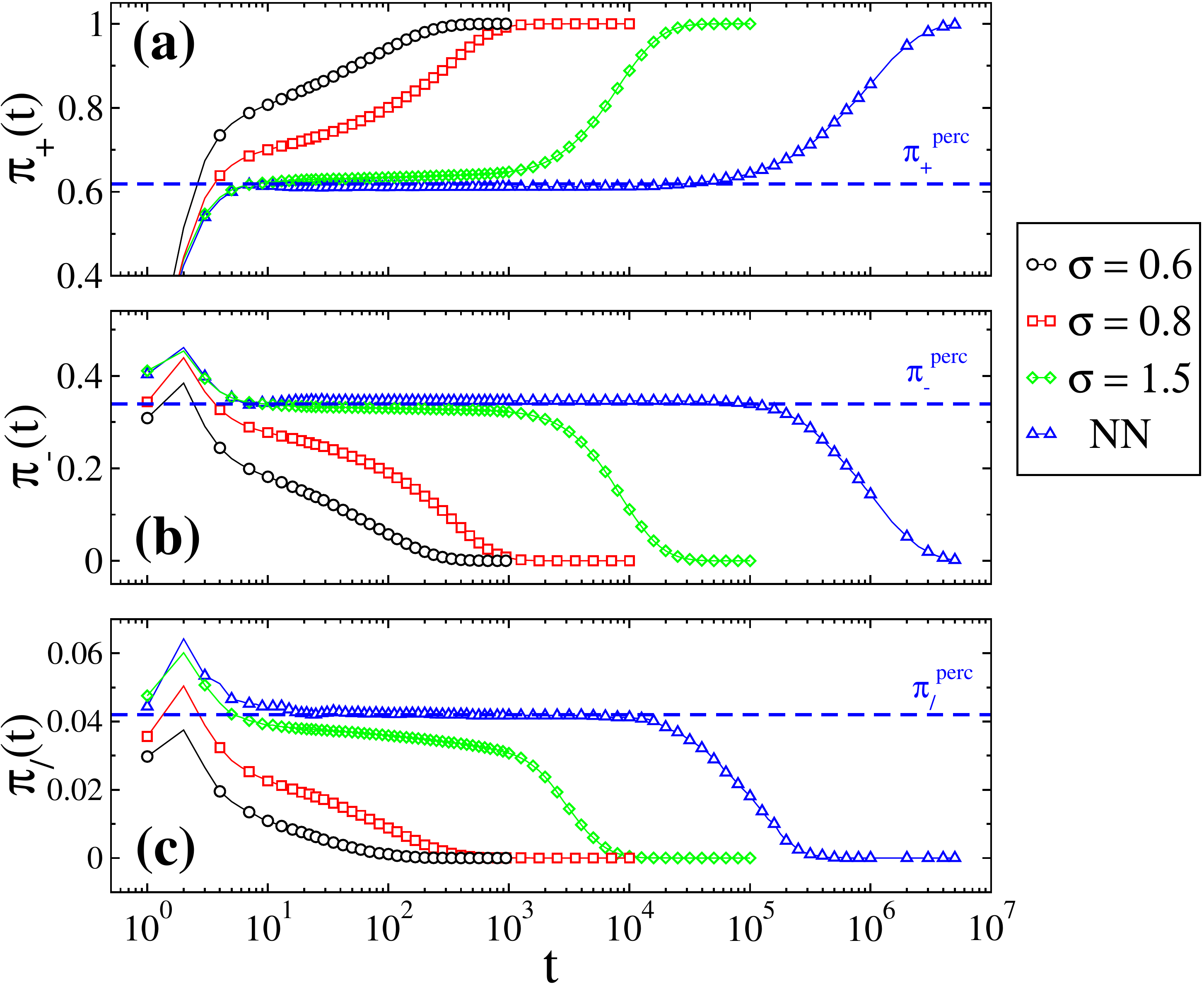}
	\caption{The crossing probabilities (a) $\pi_+(t)$, (b) $\pi_-(t)$, and (c) $\pi_/(t)$ are plotted against $t$ (with
		log-linear scales) for the NN model and for the long-range one with  
		$\sigma=0.6$, $\sigma=0.8$ and $\sigma=1.5$ (see key), for a quench to $T=0.3~T_c$ of a system of linear size
		$L=128$. The horizontal blue dashed lines represent the percolation values $\pi_+^{perc}$, $\pi_-^{perc}$, $\pi_/^{perc}$.}
	\label{figS6}
\end{figure}

Let us now shift our focus to the nature of relaxation at non-zero $T$ quenches. With NN interactions and quench to $T=0$ the configuration with vertical/horizontal stripes near $t_m$ are blocked states, but for non-zero $T$, such configurations will be eventually converted to a completely ordered state by means of activated processes. This raises the natural question of how the presence of thermal fluctuations affects the above discussed interplay of long-range interaction and system size. To gain better understanding, we perform numerical simulations for nonzero $T$ quenches and different $\sigma$. First of all, we show in Fig.~\ref{figS6} a comparison between the crossing probabilities $\pi_x(t)$ in the NN model and the long-range model, for a quench to final temperature $T=0.3~T_c$ and the same system size $L=128$. Here one sees that for NN, the system gets trapped for a long time in the striped metastable states and only at very long times (of order $10^5$), the $\pi_x$ move from the percolation values and converge to the values $\pi_-=\pi_/=0$, $\pi_+=1$ of the ordered state. Instead, with long-range interactions, the system orders at earlier times. As is clear in the figure, this time required to reach the ordered state decreases with stronger long-range interaction (small $\sigma$).

Furthermore, the time required to reach the asymptotic state $t_m$ depends on the quench temperature as well as on the system size. In the NN model, one finds~\cite{PhysRevE.65.016119} $t_m\sim L^3 t_{act}$, where $t_{act}$ is the Arrhenius time $\sim {\rm e}^{4J/T}$ the time required to create a dent on the flat stripe via the activated process, and prefactor $L^3$ comes from the diffusive processes incited after the creation of a dent. With long-range coupling, this scaling argument modifies in two respects. First, the extended interactions add a drift to the diffusive processes. Since an order $L^2$ of spins must be flipped to align all the spins, one expects $t_m\propto L^2$, instead of $t_m\propto L^3$ for NN. Another difference concerns the temperature dependence, which for NN was easily estimated from $t_{act}$ since an activated process was needed to create the dent. As discussed in Ref.~\cite{PhysRevE.103.012108}, counting the number of activated processes with long-range is a difficult task, and therefore, we cannot as easily predict the temperature dependence of $t_m$. We expect $t_m\sim L^2 f(T,\sigma)$, where $f(T,\sigma)$ is an unknown monotonically decreasing function of temperature at fixed value of $\sigma$.

\begin{figure}[t]
	\centering
	\includegraphics[width=0.48\textwidth]{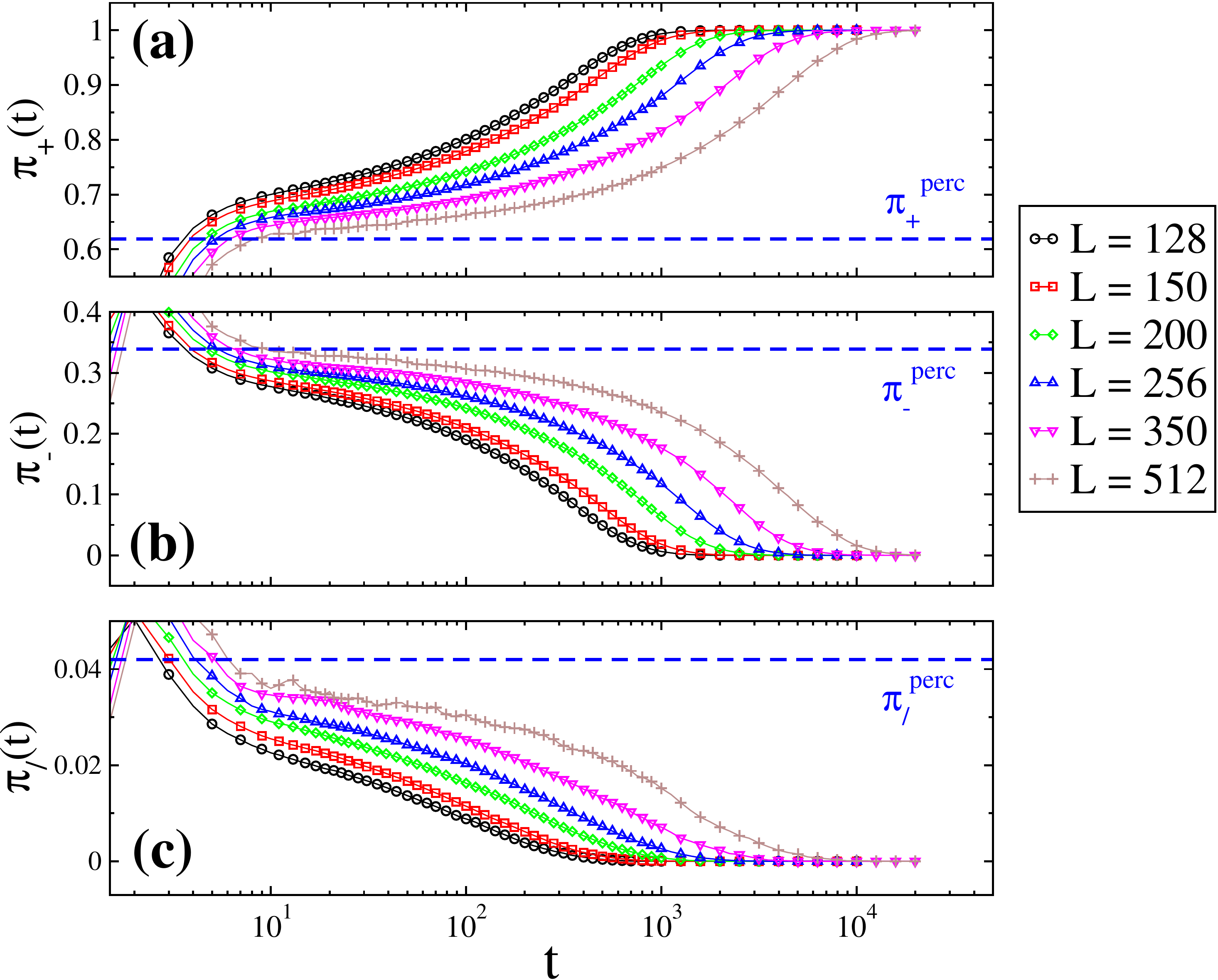}
	\caption{The crossing probabilities (a) $\pi_+(t)$, (b) $\pi_-(t)$, and (c) $\pi_/(t)$ are plotted against $t$ (with log-linear scales) for a quench to $T=0.3~T_c$ with $\sigma=0.8$ for systems of different sizes $L$, see key. The horizontal blue dashed lines represent the percolation values $\pi_+^{perc}$, $\pi_-^{perc}$, $\pi_/^{perc}$.}
	\label{figS8}
\end{figure}

\begin{figure}[t]
	\centering
	\includegraphics[width=0.48\textwidth]{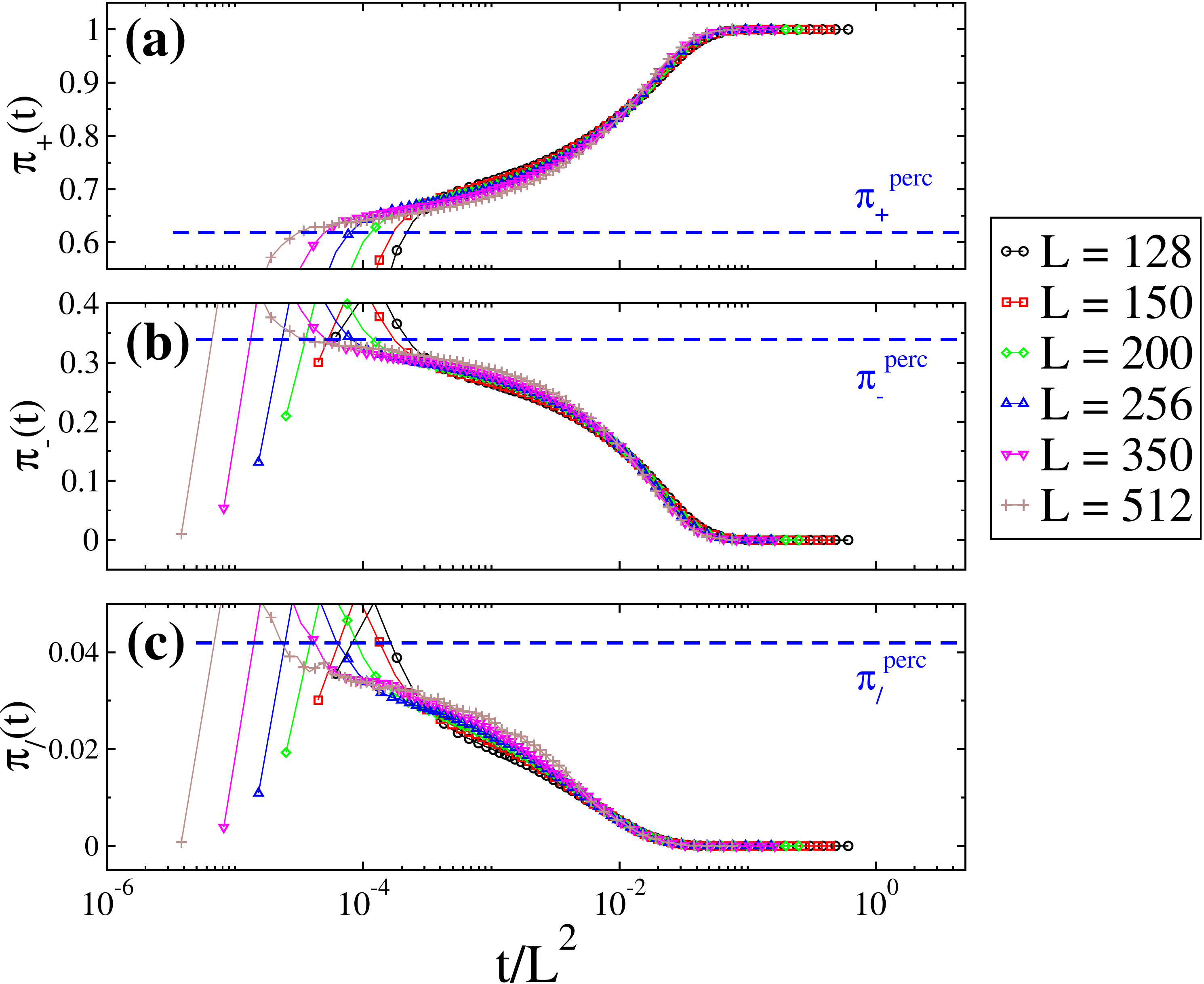}
	\caption{The crossing probabilities (a) $\pi_+(t)$, (b) $\pi _-(t)$, and (c) $\pi_/(t)$ are plotted against $t/L^2$ (with log-linear scales) for same data in Fig.~\ref{figS8}.}
	\label{fig7}
\end{figure}

In Fig.~\ref{figS8}, we plot the crossing probabilities $\pi_x(t)$ against $t$ for $\sigma = 0.8$, for systems of different sizes $L$ quenched to $T=0.3~T_c$. The figure shows that, similar to zero temperature quench, $\pi_x(t)$ first reach a pre-asymptotic plateau at height $\pi_x^{early}$. With the increase in $L$, $\pi_x^{early}$ approach the percolation values $\pi_x^{perc}$. Asymptotically, the system of any $L$ reaches the ordered state ($\pi_-=\pi_/=0$, $\pi_+=1$). However, the required time $t_m$ is increased with $L$. To testify the above proposed scaling $t_m\propto L^2$, we plot in Fig.~\ref{fig7} the $\pi_x(t)$ against the scaling variable $t/L^2$ for data in Fig.~\ref{figS8}. One can see a good collapse of data for whole evolution of $\pi_x(t)$ (after a microscopic time); indicating that $t_m \propto L^2$. Similar behavior is obtained for $\sigma = 0.6$.

\begin{figure}[t]
	\centering
	\includegraphics[width=0.48\textwidth]{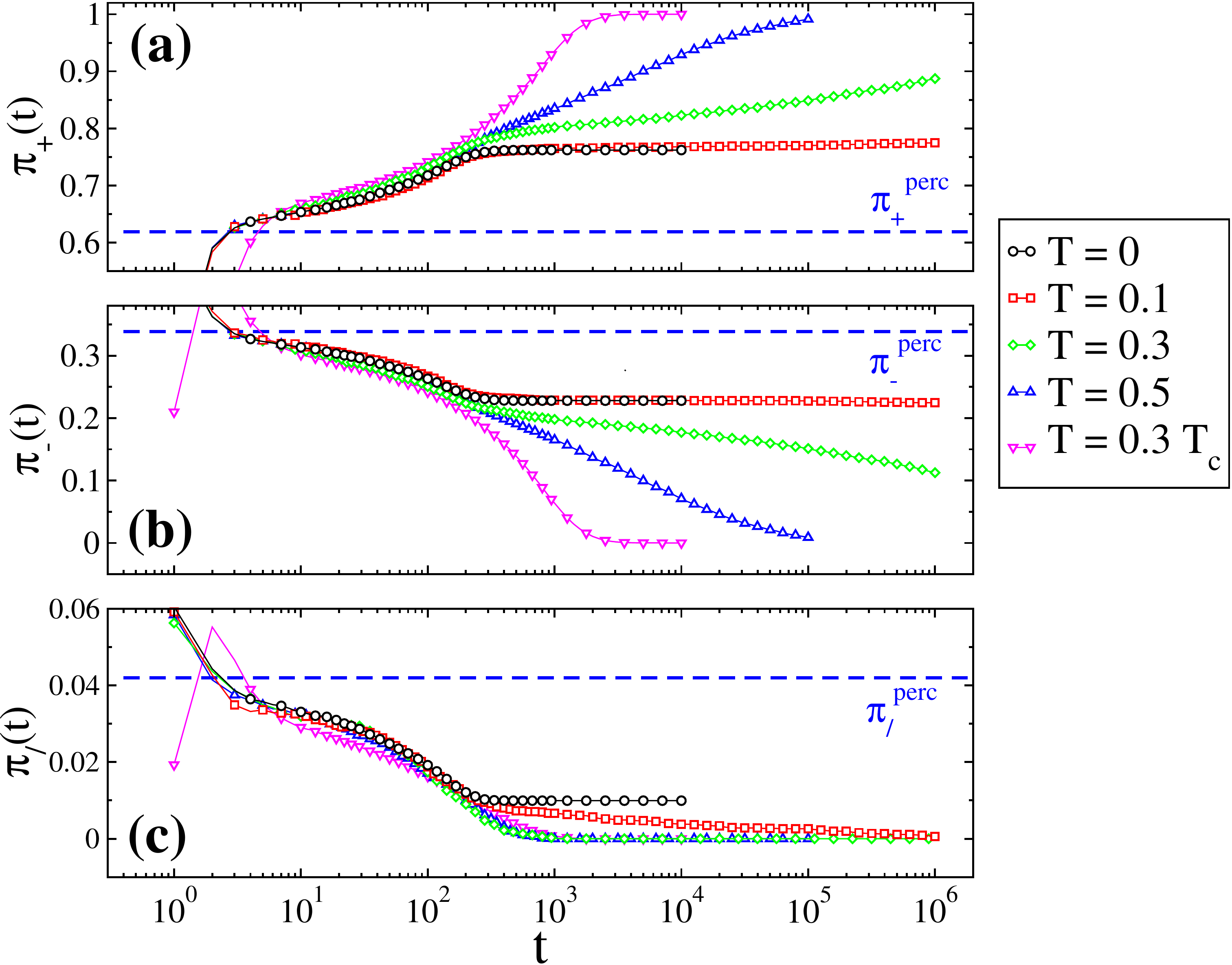}
	\caption{The crossing probabilities (a) $\pi_+(t)$, (b) $\pi _-(t)$, and (c) $\pi_/(t)$ are plotted against $t$ (with log-linear scales) for the long-range system with $L=200$ and $\sigma=0.8$ quenched to different final temperatures (see key). The horizontal blue dashed lines represent the values of $\pi_+^{perc}$, $\pi_-^{perc}$, $\pi_/^{perc}$.}
	\label{figS7}
\end{figure}

Further, to understand the temperature-dependence of equilibration time $t_m$, we plot the crossing probabilities $\pi_x(t)$ against $t$ in Fig.~\ref{figS7} for $\sigma=0.8$ and $L=200$, for different quench temperatures. We see that upon lowering the temperature, the system relaxes to the ordered state at larger and larger times. This relaxation at low temperatures occurs via two-step behavior. Initially, the system behaves as in zero temperature case, and only at large time scales it approaches the ordered state. Instead, at higher temperatures, the relaxation process is faster due to thermal agitation.

\section{Conclusion}
\label{S5}

This paper addresses the asymptotic states of two-dimensional Ising ferromagnets with power-law long-range interactions, after a quench below the critical temperature. This is the first systematic study of this problem with long-range interactions. For zero temperature quench we found that, during the early stage of coarsening, a \textit{critical percolation} structure is formed in the presence of long-range interactions, which spans across the system and determines the fate of the final states. Due to extended interactions, this structure does not always pin the system for longer times and depends on the interplay between the system size $L$ and the strength of the interaction. For finite $L$, the extended interactions convert a fraction of the percolation structures spanning in one direction to the one with spanning in both directions. This leads to the suppression of striped \textit{metastable} states. However, in the thermodynamic limit $L \rightarrow \infty$, the same percolation structure remains in the system up to the equilibration time $t_m$, and the striped states occur with the same probabilities as in the short-range model~\cite{BarKraRed09,PhysRevLett.109.195702}.

For non-zero temperature quenches, the present system always reaches the ordered state by means of activated processes. Again, the system size $L$ and the long-range interaction play a major role. The relaxation to the ordered state is faster for stronger interaction strength, i.e., small $\sigma$. With the increase in $L$, the convergence to ordered state is delayed to larger time, which scales as $t_m \propto L^2$.

\begin{acknowledgments}
We thank one of the referees for suggesting the scaling in Eq.~\eqref{newscale}.
\end{acknowledgments}

\appendix

\begin{figure}[t!]
	\centering
	\includegraphics[width=0.48\textwidth]{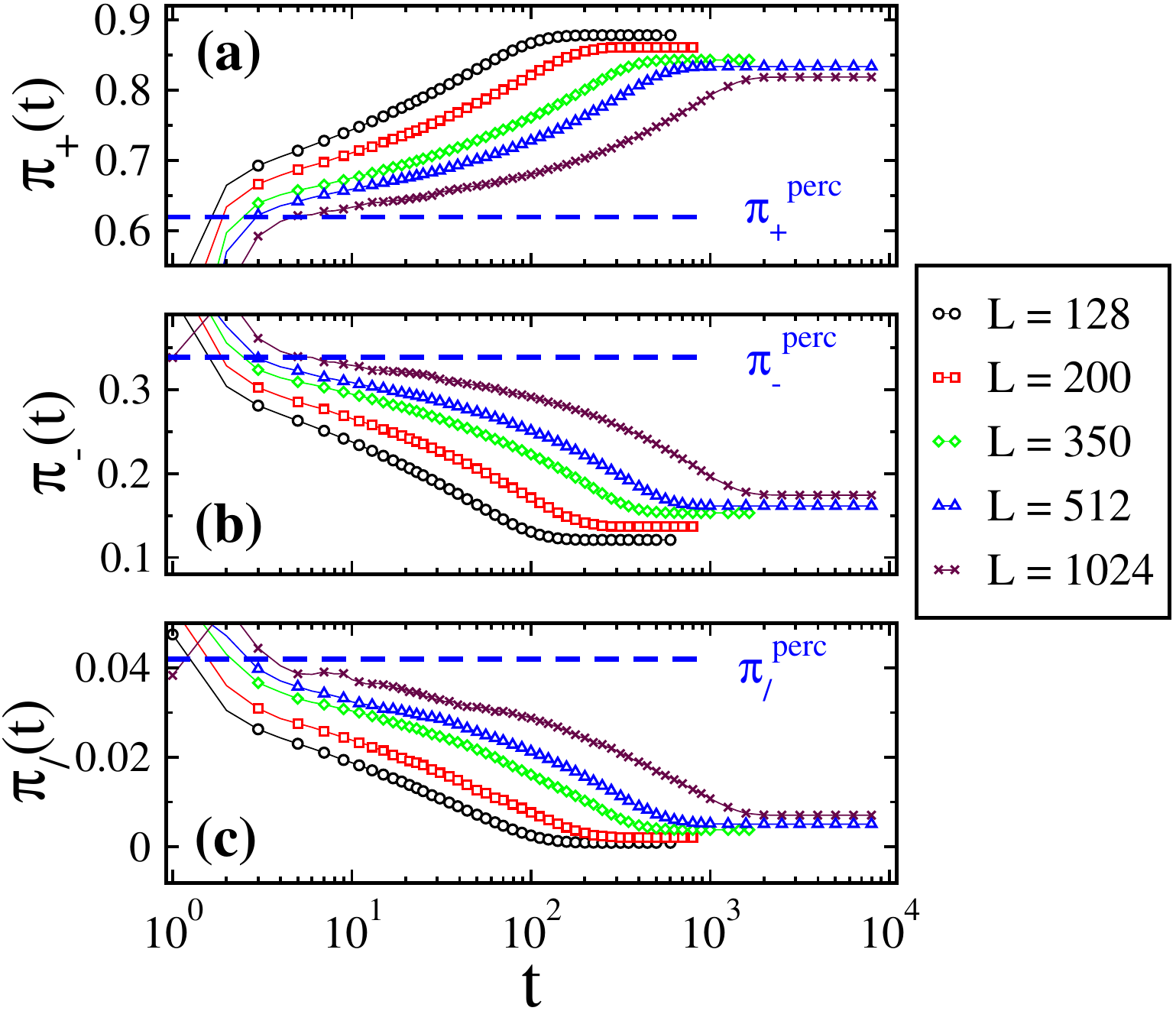}
	\caption{The crossing probabilities (a) $\pi_+(t)$, (b) $\pi_-(t)$, and (c) $\pi_/(t)$ are plotted against $t$ (with log-linear scales) for a quench to $T=0$ with $\sigma=0.6$ for systems of different sizes $L$, see key. The horizontal blue dashed lines represent the values of $\pi_+^{perc}$, $\pi_-^{perc}$, $\pi_/^{perc}$.}
	\label{figS1}
\end{figure}

\begin{figure}[t!]
	\centering
	\includegraphics[width=0.48\textwidth]{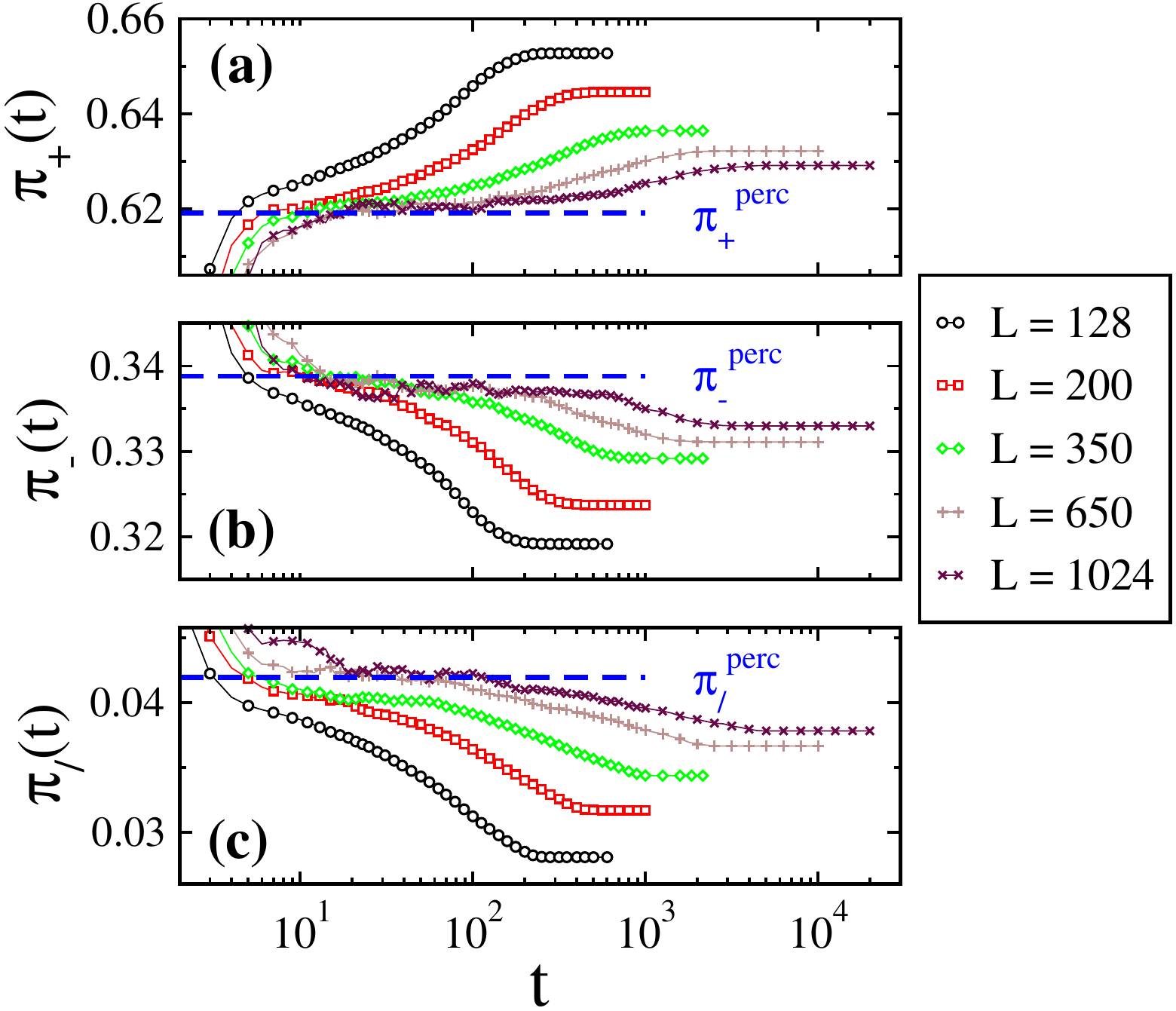}
	\caption{The crossing probabilities (a) $\pi_+(t)$, (b) $\pi_-(t)$, and (c) $\pi_/(t)$ are plotted against $t$ (with log-linear scales) for a quench to $T=0$ with $\sigma=1.5$ for systems of different sizes $L$, see key. The horizontal blue dashed lines represent the values of $\pi_+^{perc}$, $\pi_-^{perc}$, $\pi_/^{perc}$.}
	\label{figS2}
\end{figure}

\section{System-size dependence of crossing probabilities $\pi_x(t)$ for $\sigma = 0.6, 1.5$}
\label{A1}

In Fig.~\ref{figS1}, the crossing probabilities $\pi_x(t)$ are plotted against $t$ for $\sigma = 0.6$, for different system sizes $L$ quenched to $T=0$. It is clear that, with an increase in system size, the pre-asymptotic plateau of height $\pi _x^{early}$ settles to $\pi_x^{perc}$. Apart from that, the saturation values $\pi_x^{late}$ also reduce (and approach $\pi_x^{perc}$) with increase in $L$.

Fig.~\ref{figS2} shows plots of $\pi_x(t)$ vs. $t$ for $\sigma = 1.5$, for different system sizes $L$ quenched to $T=0$. As for $\sigma=0.6$ (see above) and $\sigma = 0.8$ (see main text), with increase in $L$, both $\pi _x^{early}$ and $\pi_x^{late}$ approach the percolation values $\pi_x^{perc}$. Notice that the range of time where $\pi_x(t)$ stays near $\pi_x^{perc}$ (especially for higher $L$) is larger for higher $\sigma$.

\section{Plots of $[\pi_x(t)-\pi_x^{perc}]/\Delta _x(L)$ vs. $R(t)/L$ for $\sigma = 0.6, 1.5$}
\label{A2}

\begin{figure}[t!]
	\centering
	\includegraphics[width=0.48\textwidth]{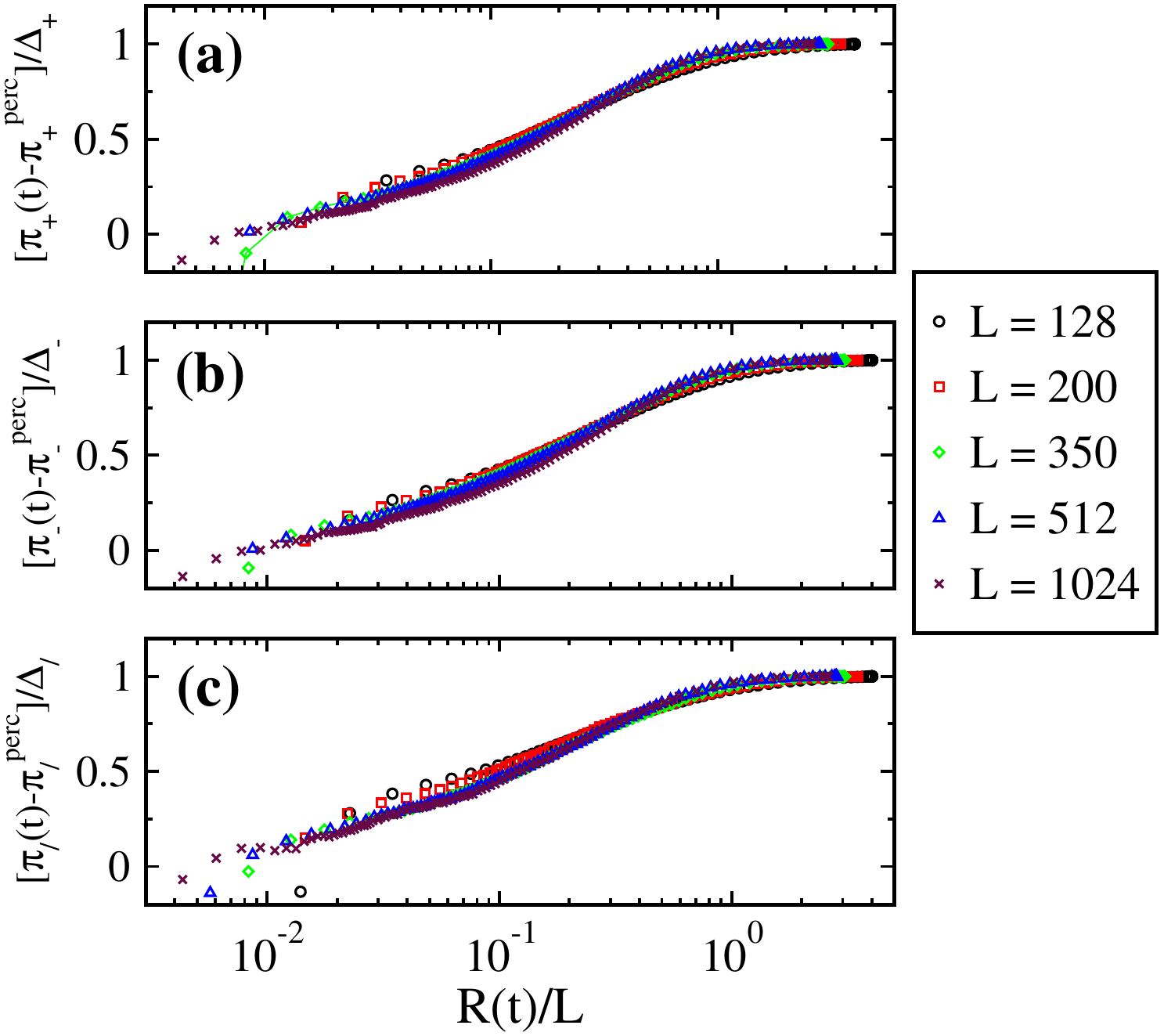}
	\caption{The quantities (a) $[\pi_+(t)-\pi_+^{perc}]/\Delta _+(L)$, (b) $[\pi_-(t)-\pi_-^{perc}]/\Delta _-(L)$, and (c) $[\pi_/(t)-\pi_/^{perc}]/\Delta _/(L)$ are plotted against $R(t)/L$ (with log-linear scales) for a quench with $\sigma=0.6$ to $T=0$ for systems of different sizes $L$.}
	\label{figS3}
\end{figure}

\begin{figure}[t!]
	\centering
	\includegraphics[width=0.48\textwidth]{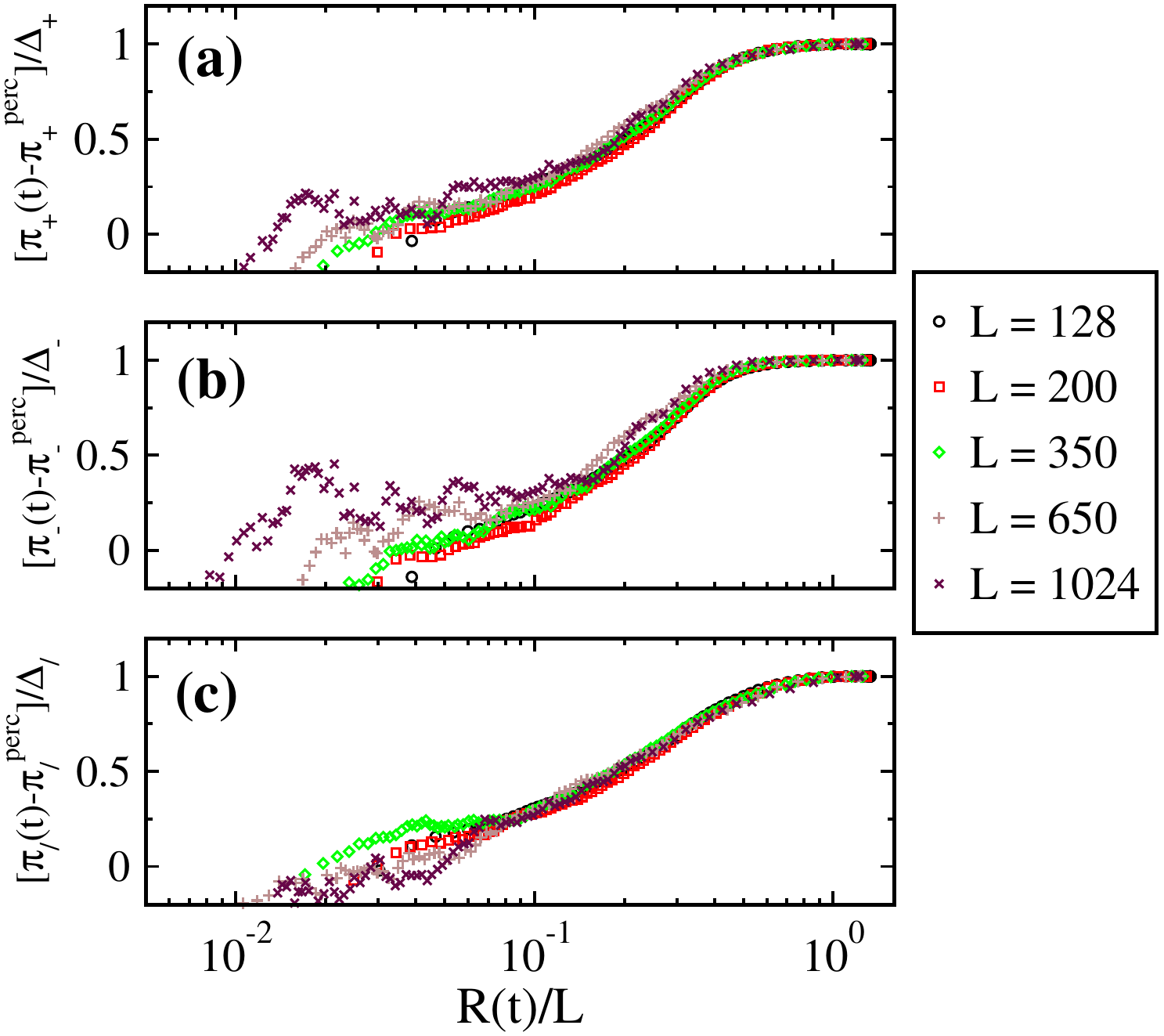}
	\caption{The quantities (a) $[\pi_+(t)-\pi_+^{perc}]/\Delta _+(L)$, (b) $[\pi_-(t)-\pi_-^{perc}]/\Delta _-(L)$, and (c) $[\pi_/(t)-\pi_/^{perc}]/\Delta _/(L)$ are plotted against $R(t)/L$ (with log-linear scales) for a quench with $\sigma=1.5$ to $T=0$ for systems of different sizes $L$.}
	\label{figS4}
\end{figure}

In Figs.~\ref{figS3} and~\ref{figS4}, we plot $[\pi_x(t)-\pi_x^{perc}]/\Delta _x(L)$ vs. $R(t)/L$ for $\sigma=0.6$ and $\sigma=1.5$, respectively. A good data collapse, irrespective of the crossing type, is observed for both $\sigma=0.6$ and $\sigma=1.5$.

\section{Pair-connectedness}
\label{A3}

\begin{figure}[b!]
	\includegraphics[width=0.48\textwidth]{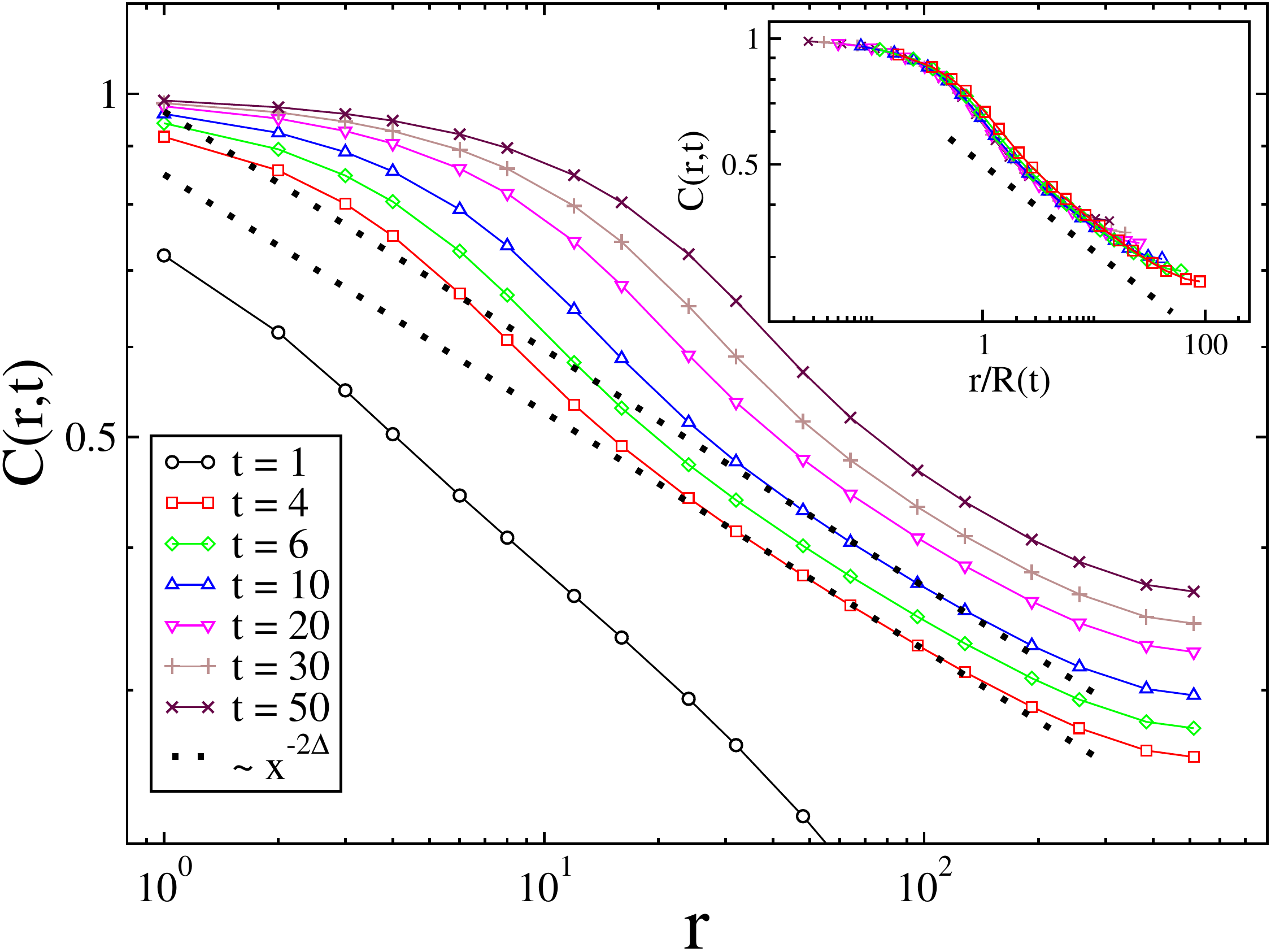}
	\caption{The pair-connectedness function $C(r,t)$ is plotted against $r$ with double logarithmic scales, at different times (see key) after the quench to $T=0$ of a long-range system of linear size $L=1024$ with $\sigma =0.8$. The dotted lines are the power laws given by Eq.~\eqref{perc_pc}, with the exponent $2\Delta=5/24$ of critical percolation. In the inset, the same data are plotted against scaling variable $r/R(t)$.}
	\label{figS5}
\end{figure}

Given a discrete lattice, the pair-connectedness function gives the probability that two points at a distance $r$ are part of the same cluster. In two-dimensional ($d=2$) random percolation model at percolation threshold $p=p_c$, it is given as~\cite{stauffer2018introduction}
\begin{equation}
C_{perc}(r,r_0)=\left[ \frac{r}{r_0} \right]^{-2\Delta},
\label{perc_pc}
\end{equation}
with $\Delta = 5/48$ being a critical exponent. Here $r_0$ is the lattice spacing. The above relation holds for $r\gg r_0$.

In a spin system, the pair connectedness function at time $t$ can be calculated from the connected domains of positive and negative spins as follows:
\begin{equation}
C(r,t)=\frac{1}{4L^2} \sum_i \sum_{i_r} \langle  \delta_{s_i, s_{i_r}}   \rangle ,
\label{spin_pc}
\end{equation}
where $\langle \cdots \rangle $ is a non-equilibrium average, $L$ is the linear size of the $2d$ square lattice, and the index $i_r$ indicates the local neighboring spins at distance $r$ from $s_i$. The quantity $\delta_{s_i, s_{i_r}}=1$ if the two spins belong to the same cluster, $\delta_{s_i, s_{i_r}}=0$ otherwise. $C(r,t)$ is useful to recognize the geometry of critical percolation in a coarsening spin system by replacing $r_0 \rightarrow R(t)$ in Eq.~\eqref{perc_pc}, where $R(t)$ is the characteristic length-scale in the coarsening system. Thus, one has
\begin{equation}
C(r,t) \sim \left[ \frac{r}{R(t)} \right]^{-2\Delta},~~~~~~\mbox{for} \quad r\gg R(t).
\label{scalpercpc}
\end{equation}
As time $t \gg t_p$, the system can be considered a lattice at critical percolation with lattice spacing $R(t)$.

In Fig.~\ref{figS5}, we calculate the pair-connectedness function $C(r,t)$ for $\sigma=0.8$ and $L=1024$, after a quench to $T=0$. In the main frame, $C(r,t)$ is plotted against distance $r$ for different times. Initially ($t\simeq 1$), $C(r,t)$ deviates from the power-law behavior of critical percolation~\eqref{perc_pc}. This is because a critical percolation structure has not yet emerged in the system. At later times, for $r$ larger than a typical value that increases in time ( i.e., $R(t)$, see Ref. \cite{CorCugInsPic17} for more details), the curves at different times agree with Eq.~\eqref{perc_pc}. Further, because of finite-size effects, this agreement does not hold till large distances. This behavior is analogous to what observed with the nearest-neighbor (NN) case~\cite{Blanchard_2017,CorCugInsPic17}. In the inset of the same figure, $C(r,t)$ is plotted against the scaling variable $r/R(t)$ for the same data in the main frame. The obtained collapse stresses the validity of scaling relation~\eqref{scalpercpc}, confirming the critical percolation geometry of the spin clusters in the long-range model. Again, the collapse is not observed at large $r$ due to the finite-size effect.

\bibliography{perclr}

\end{document}